\documentclass[aps,prc,amsmath,amssymb,twocolumn,showpacs]{revtex4-1}
\usepackage{graphicx,amsmath,amssymb,bm,soul,cancel}
\usepackage[usenames]{color}
\usepackage{natbib}
\usepackage{graphicx}
\usepackage{amsmath}
\usepackage{hyperref}
\usepackage{tikz}

\allowdisplaybreaks

\DeclareMathOperator{\sech}{sech}  % PT potential

\begin{document} 
    \title{Machine-learning approach to finite-size effects in systems with
    strongly interacting fermions}
    \author{Nawar Ismail}
    \affiliation{Department of Physics, University of Guelph, Guelph, ON N1G 2W1, Canada}
    \author{Alexandros Gezerlis}
    \affiliation{Department of Physics, University of Guelph, Guelph, ON N1G 2W1, Canada}
    % \date{\today}
    \begin{abstract}
        We investigate the applicability of machine learning techniques in studying the finite-size effects associated with many-body physics.
        These techniques have an emerging presence in many-body theory as they have been used for interpolations, extrapolations, and in modeling wavefunctions. 
        We will resolve several issues associated with machine learning and many-body calculations such as small datasets, outliers, and discontinuities, for the purpose of extrapolating finite calculations to macroscopic scales.
        We carry out a systematic investigation of two related systems by developing metrics that aim to avoid spurious effects and capture desired features.
        This work uses neural networks to extrapolate the Unitary Gas to the thermodynamic limit at zero-range, which is otherwise difficult to reach. 
        The effective mass of strongly interacting neutron matter is also studied and makes use of the non-interacting problem to resolve discontinuous predictions. For this investigation, we also carried out new Auxiliary Field Diffusion Monte Carlo (AFDMC) calculations for a variety of densities and particle numbers.
        Ultimately, we demonstrate an effective utility for neural networks in this context.
    \end{abstract}
    \maketitle 

    \section{Introduction}
        In recent years machine learning (ML) techniques have established themselves in quantum 
        many-body theory as a set of essential and promising tools for solving a diverse range of novel and existing problems. In the study of nuclei, extrapolations from the no-core shell model along with coupled-cluster calculations have been carried out to determine nuclear masses and radii \cite{jiang2019extrapolation, Negoita_2019}. Machine learning has also been applied to improve computational efficiency \cite{huang_2017, liu_2017, shen2018selflearning, Nelson_2019, Torlai_2018, Carrasquilla_2019} and to develop computationally feasible models, like those for wavefunctions \cite{Lasseri_2020, Carleo_2017, Gao_2017}. In general, the utility provided by these machine learning models arises from their ability to capture underlying dependencies, allowing for data to be interpolated and extrapolated. Despite the ability of the networks to generalize, special attention must often be made to the representation of the data through, for example, feature engineering, to avoid misleading extrapolations \cite{pastore2020extrapolating}. 

        The \textit{ab initio} study of strongly interacting systems is often limited by prohibitive computational costs, which constrain calculations to finite domains, typically in terms of particle number. In the study of
        the unitary gas (UG) and of neutron matter (NM), these limitations arise when solving the quantum many-body problem and prevent certain calculations from significantly exceeding $\approx 100$ particles. A large focus is appropriately placed on investigating the difference between these finite calculations and the infinite system, which are known as the finite-size effects (FSE) \cite{Kwee:2008, giorgini:2008, Carlson_2012, jiang2019extrapolation, gandolfi:2009, gezerlis:2013, giorgini:2008}. For systems with a fixed density, the infinite system is referred to as the thermodynamic limit (TL) and best describes macroscopic matter. In general, these deviations tend to diminish as the number of particles rises, although the behavior is often neither smooth nor monotonic. Typically, careful analysis is required to determine which finite particle number best matches the TL. The ability to interpolate these dependencies can provide great insight into the limiting behavior. Naturally, this type of problem falls in the domain of machine learning, and so we aim to apply those techniques to study these finite-size effects.

        The Neural Network (NN) has become a ubiquitous machine learning model due to its effectiveness in a variety of problems. Such an approach takes in a set of inputs and, through some scheme, propagates it through the layers of the network until an output is achieved. The particular scheme used to derive an output depends on the organization and structure of the network, which is known as its architecture. The internal weights used in the model are typically organized into so-called hidden layers, which vary depending on the type of network. Feed-forward neural networks (FFNN) have been used extensively in nuclear physics \cite{pastore2020extrapolating, Negoita_2019, jiang2019extrapolation, Lasseri_2020, adams2021variational}. Other effective neural networks are the Boltzmann Machine (BM) \cite{Gao_2017, Carleo_2017, huang_2017, Rrapaj_2021} and recursive neural networks (RNN) \cite{Guest_2018}. Sometimes tailored architectures are required or are beneficial for solving particular problems \cite{Raghavan_2021}.

        As noted, an  FFNN is organized into a set of layers, where the first layer receives input values that are propagated through the hidden layers until an output is produced \cite{jiang2019extrapolation, keeble2019machine}. Each layer is comprised of nodes that are connected to each node in the subsequent layer. These connections are viewed as weights since the value of a node, $y$, depends on the previous layer $\boldsymbol{x}$ through a weighted sum:
        \begin{equation}
            y = f(\boldsymbol{w}\cdot\boldsymbol{x} + b),
        \end{equation}
        where $\boldsymbol{w}$ and $b$ are free parameters that can be tuned, and $f$ is an activation function that can be used to introduce non-linearities to the model. The process is iterated until the final output(s) are generated. Multiple hidden layers provide a hierarchical structure to the network that can capture more complex features, however training these networks can require more time to train, and care to avoid problems such as gradient vanishing. 
        However, one of the key properties of the neural network is the Universal Approximation Theorem, which states that a single hidden layer is sufficient to approximate any continuous function, provided the layer has a large enough number of neurons \cite{pastore2020extrapolating, jiang2019extrapolation}. Of course, it does not suggest the ideal number of nodes in a layer or what the values of the internal weights should be, and these must instead be found during an optimization procedure known as training. During this optimization procedure, the small datasets often encountered in nuclear physics can pose challenging issues. 

        To identify and resolve difficulties that may arise in this context, we will be applying machine learning techniques to two related systems. In the case of the unitary gas, we encounter outliers and determine their impact by studying how network predictions vary based on how much emphasis is placed on them during training. To mitigate the negative effects of a small dataset, the technique of data augmentation is used to increase the dataset size without significant additional cost \cite{jiang2019extrapolation}. Other issues arise during the study of the effective mass of strongly interacting neutron matter. We performed additional energy evaluations using Auxiliary Field Diffusion Monte Carlo (AFDMC) to generate a dataset, which is provided as supplementary material in Ref.~\cite{supple}. Due to discontinuities in the dataset, the networks initially provided spurious predictions, which could be corrected by using the non-interacting problem. After systematically studying these effects, TL predictions were arrived at for both problems.

        \section{Methods}
            \subsection{Potentials}
Neutron matter plays an important role in understanding neutron-rich nuclei and neutron stars (NS)\cite{gandolfi:2015, Buraczynski_2020} and it can be described by the following Hamiltonian:
                \begin{align}\label{eq:hamiltonian}
                    \hat H = -\frac{\hbar^2}{2m}\sum_i \nabla^2_i + \sum_{i<j}v_{ij} + \sum_{i<j<k} v_{ijk},
                \end{align}
                which considers the kinetic energy, a two-body interaction, and a three-body interaction. 
                Higher-order considerations,
like four-body interactions, have an effect on the total energy of the many-body system
that is an order of magnitude smaller than that arising from three-nucleon interactions~\cite{Tews:2013};
thus, we may safely ignore them. Many forms of potential exist that capture varying aspects and levels of detail from the underlying nuclear interaction; typically either phenomenological \cite{carlson:2003, gandolfi:2009, gezerlis:2010, gandolfi:2012, baldo:2012} or 
                effective-field theory potentials \cite{Hebeler_2010, gezerlis:2013, machleidt:2011, Hagen_2014, gezerlis:2013, Carbone_2014, Roggero_2014, PhysRevLett.113.182503, PhysRevC.89.061301, PhysRevC.93.024305, Piarulli_2018, Lonardoni_2018} are used.

                In this work, high-precision calculations are carried out for the phenomenological two-body Argonne v8' (AV8') and three-body Urbana IX (UIX) potentials. The Argonne potential is comprised of spin, tensor, spin-orbit, and isospin operators with radial dependencies that are tuned to a large body of neutron-proton scattering data resulting in high-quality fits \cite{fantoni:2000, carlson:2014, gandolfi:2009, gandolfi:2015, baldo:2012}. The Urbana potential is similarly fit to light nuclei and nuclear matter \cite{gandolfi:2009, pudliner:1997}. 
                
                In the low-density regime of neutron star crusts, the potential can be effectively parameterized by the scattering length and effective range, which greatly simplifies many considerations \cite{lacroix:2017, bulgac2010unitary}. Given the scale-independence of the UG, the fine details of the interatomic potential have little impact on observables \cite{forbes2012effectiverange, Dawkins_2020}. This universal behavior allows us to group energy calculations from multiple potentials. Here we employ the diffusion Monte Carlo (DMC) results reported on in Ref.~\cite{forbes2012effectiverange} as the input; these include calculations from both the modified-P{\"o}schl-Teller potential and the double Gaussian potential which are given by:
                \begin{equation}
                		v_{PT}(r) = 4\mu_{PT}^2 \sech^2(\mu_{PT} r)
                \end{equation}
                and 
                \begin{equation}
                    v_{2G}(r) = 3.144 \mu_{2G}^2 \left(
                		e^{-\mu_{2G}^2 r^2/4} - 4e^{-\mu_{2G}^2 r^2}
                	\right),
                \end{equation}
                respectively, where $\mu_{PT} = 2/r_e$, $\mu_{2G}=3.952/r_e$, $r_e$ is the effective range, and $r$ is the distance between two particles. These potentials are essentially zero-range
                two-body s-wave interactions; there is no three-body interaction \cite{forbes2012effectiverange, castin2011unitary}.
                    
            \subsection{Quantum Monte Carlo}
                Energy calculations are carried-out by using Eq.~(\ref{eq:hamiltonian}) for the ground-state energy through the use of quantum Monte Carlo (QMC) algorithms \cite{PhysRevLett.113.182503, gezerlis:2013, gandolfi:2012, baldo:2012, buraczynski:2016}. Auxiliary Diffusion Monte Carlo (AFDMC) is a specialization of QMC that allows for high-precision calculations for up to about 100 particles \cite{sarsa2003neutron, pilati:2014, reynolds:1982}. The computational complexity of this algorithm (with respect to particle number) is largely dominated by wave function evaluations and limits us to studying finite systems \cite{Buraczynski_2020, fantoni:2000}. A periodic boundary condition is applied which acts to approximate the macroscopic scale. 
This boundary condition is applied on a cubic box of length $L$ and results in the discretization of the available momentum states such that the allowed wave-vectors are given by $\boldsymbol{k} = (2\pi/L) (n_x, n_y, n_z)$, and the $n$'s are restricted to integers. It is easy to see that there exist many combinations of 
$n$'s that lead to the same wave number/energy.
                At finite $N$, these calculations deviate from the (otherwise unknown) TL values. These deviations are referred to as the Finite-Size Effects (FSE) and a systematic study is required to make claims about macroscopic neutron matter \cite{Kwee:2008, carlson:2012, gandolfi:2009, giorgini:2008}.
                As in most QMC works, we here employ periodic boundary conditions; a more general
scheme, twisted boundary conditions, leads to distinct behavior which would have to be separately modelled~\cite{Riz:2020,Palkanoglou:2021}.

            \subsection{Machine Learning}
In a task known as \textit{supervised learning}, machine learning algorithms learn to make predictions from datasets that contain labelled data \cite{Athanassopoulos_2004}. This training process aims to minimize the deviations from the predictions of a model and the dataset. The goal here is that, in addition to reproducing the dataset, the machine learning model is also capable of extrapolations or interpolations. This property of generalization is essential to avoid \textit{overfitting}, whereby the model fails to capture underlying features and is only capable of reproducing the given examples \cite{jiang2019extrapolation, carrasquilla2016machine, shen2018selflearning}.

                For this work, the feed-forward neural network \cite{jiang2019extrapolation, Negoita_2019, Athanassopoulos_2004} will be used to perform various regressions. Incidentally, when we employ below the term ``neural network'' we typically refer to an \textit{ensemble} of networks, generated by using random initial conditions. The internal weights of the network are tuned to minimize the distance between the predictions and the dataset. This tuning occurs during a training process that typically uses the back-propagation and gradient-descent algorithms (and its variants) \cite{jiang2019extrapolation, keeble2019machine}. Our machine-learning calculations will be carried out using the Keras Python library with a Tensorflow backend and using densely connected layers.
                
        	\begin{figure}
\begin{center}
\includegraphics[width=1.0\columnwidth,clip=]{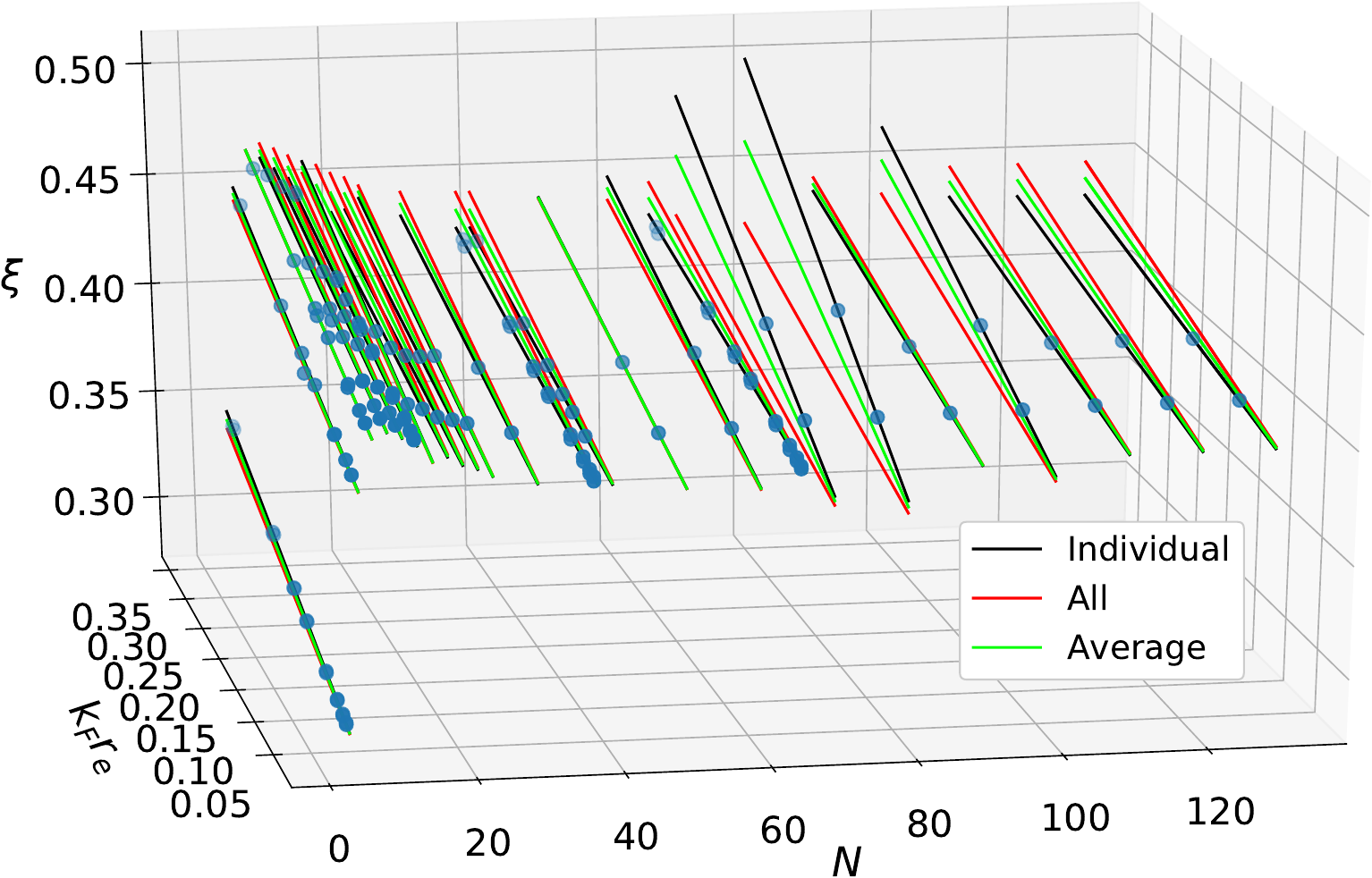}
        		\caption{
        			The Unitary Gas dataset plotted along with three types of linear fits given by Eq.~(\ref{eq:xi_individual}),~(\ref{eq:xi_all}),~and~(\ref{eq:xi_average}), which correspond to the ``individual'', ``all'', and ``average'' fits, respectively. The effective-range dependence is strongly linear while the particle-number dependence is more complex. At $N=70$ and $N=80$, the slopes of the different fits notably disagree almost exclusively at these points, which suggests that they may be spurious.
        		}
        		\label{fig:UG_dataset}
        		\end{center}
        	\end{figure}

                A dataset consists of examples. Each example contains an input and an output, which may each consist of multiple values. Networks trained on a given dataset are susceptible to \textit{overfitting}, which describes the tendency for networks to generate poor predictions for examples that are not included in the training dataset (despite having a low error on the training examples). To avoid this issue, datasets are typically divided into a \textit{training} portion and \textit{testing} portion. The training portion is used to optimize the internal weights, while the testing portion is used to measure the ability of the network to generalize since the testing dataset contains novel examples. A sophisticated version of this, known as k-fold cross-validation \cite{raschka2020model}, can be used to improve data usage during training as is done in the UG investigation.
                This technique divides the dataset into $k$ subsets; $k-1$ subsets are used during training, while 1 subset is left out as the test set to evaluate model performance. Since this test set can be any of the $k$ subsets, the results are averaged across all permutations. Meanwhile, the effective mass will require additional considerations.
                
                The optimization procedure attempts to determine optimal values for the internal weights of a model that allows it to generalize well. In addition to the internal weights of the neural network, there are a set of \textit{hyperparameters} that also impact the predictions of the trained network. These may include factors such as the size of the network, the number of training iterations (epochs), the use of a regularizer, among others. The performance of networks with different hyperparameters can be measured according to a prescribed metric, like the cross-validation error. Networks that perform well here are typically selected as those that have captured the underlying features.

    \section{Learning the Unitary Gas}
        \subsection{Dataset}
            The UG inputs we are faced with are parametrized by two parameters, the effective range $r_e$ and a finite particle number $N$ \cite{forbes2012effectiverange}. The effective range is typically expressed as a dimensionless quantity, $k_F r_e$, where $k_F$ is the Fermi momentum. Similarly, the energy $E$ is also represented as a dimensionless quantity $\xi=E/E_{FG}$, where $E_{FG}$ is the energy of the free Fermi gas.

        	The dataset shown in Fig.~\ref{fig:UG_dataset} demonstrates that the effective-range dependence is strongly linear while the $N$-dependence is more complex, as noted in Ref.~\cite{forbes2012effectiverange}. This is useful since linear fits can be used to \textit{augment} the dataset \cite{jiang2019extrapolation}, by providing additional points of data without much cost. Although higher-order fits (e.g., quadratic) may have some benefit, many of the $N$ only correspond to two data points, and so for consistency we will employ linear fits. 
        	The slopes and intercepts can be determined in multiple ways:
        	\begin{align}
        		\xi_\text{individual}(k_Fr_e;N) &= S(N)k_Fr_e + b(N)\label{eq:xi_individual}\\
        		\xi_\text{overall}(k_Fr_e;N) &= \langle S(N) \rangle k_F r_e + b(N)\label{eq:xi_all}\\
        		\xi_\text{average}(k_Fr_e;N) &= \frac{S(N) + \langle S(N) \rangle}{2} k_Fr_e + b(N)\label{eq:xi_average}
        	\end{align}
        	where $S(N)$ and $b(N)$ refer to the slope and intercept of the line that best fits through the points at a given $N$.  
The angle bracket notation denotes the average slope across all $N$.        	
Studying these different fits will provide insight into the dataset.        	
        	The individual fit captures only local effects, the overall captures only global effects, and the average attempts to capture both. These linear fits are shown in Fig.~\ref{fig:UG_dataset}. Upon comparing these, outliers at $N=70$ and $N=80$ are identified. Since we are aiming to capture large $N$, removing these outliers prematurely may be counter-productive, so determining their impact is important.
        	
        \subsection{Outliers}\label{sec:outliers}
            These outliers occur particularly at large $N$. To understand their effect, we train networks on datasets that emphasize different $N$ through the means of upsampling. Let's denote the number of data points at a given particle number in the original dataset as $C_N^\text{original}$. We can tune the balance by linearly interpolating according to:
    		\begin{equation}
    			C^\text{balanced}_N(t) = tC^\text{original}_N + (1 - t)\max(\{ C^\text{original}_i \}).
    		\end{equation}
    		where the second term picks the maximum number of data points across particle numbers.
    		There are a few key values: $t=1$ describes the original dataset, $t=0$ describes a uniform distribution, and $t=-1$ describes a \emph{reflected} distribution. The total counts here depend on $t$, which is undesired, and so the counts are normalized according to:
    		\begin{equation}\label{eq:C_N^F}
    			C^\text{fixed}_N(t) = \left\lceil \frac{F}{\sum_N C^\text{balanced}_N(t)}\right\rceil C^\text{balanced}_N(t),
    		\end{equation}
    		where the ceiling function is used to provide an integer count at least as large as $F$; this $F$ is the hyperparameter controlling the number of data points.

    		\begin{figure}
    			\centering
 \begin{center}
\includegraphics[width=1.0\columnwidth,clip=]
   			{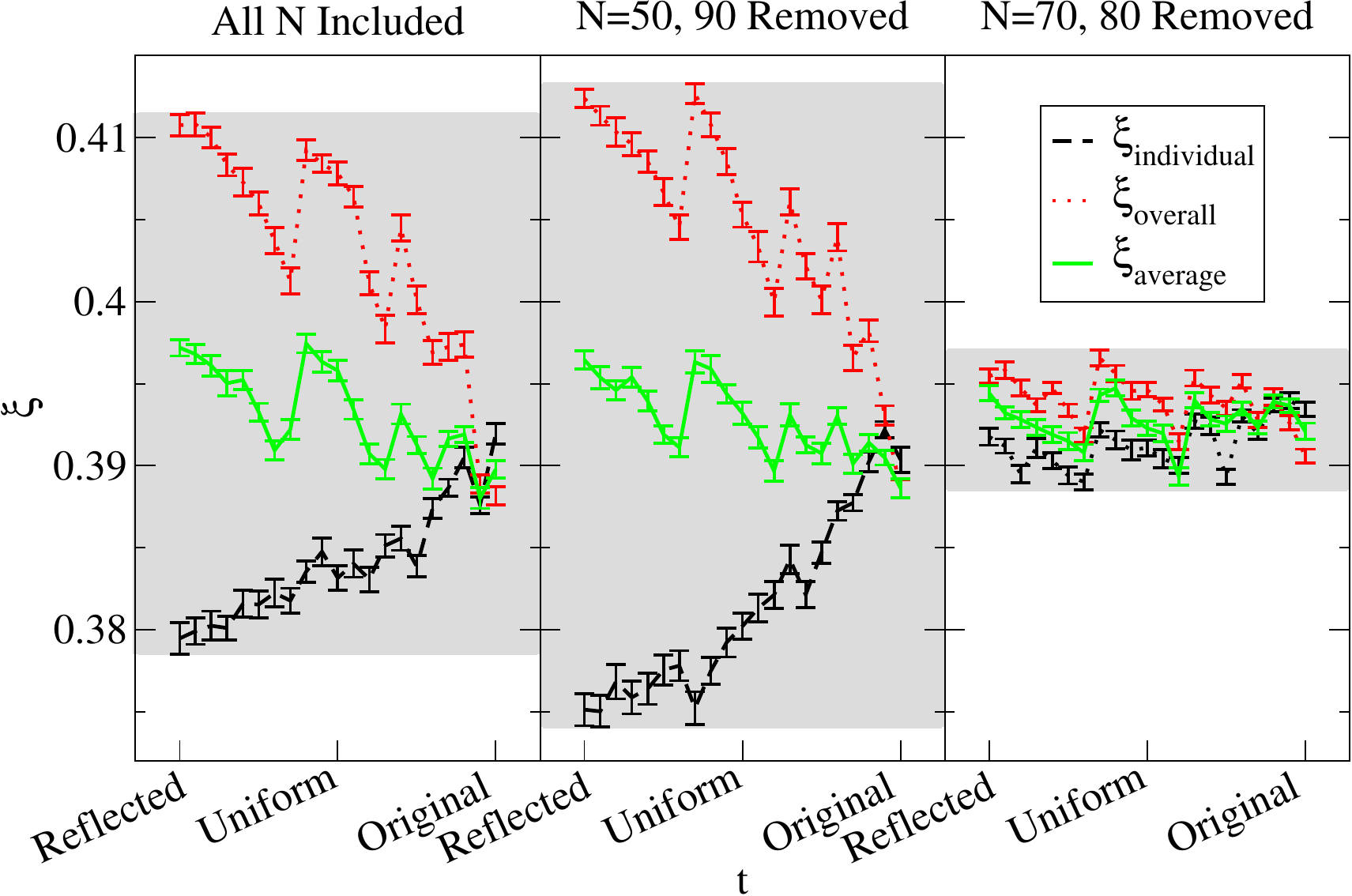}
    			\caption{  % Could say 'subplot' instead of 'panel'
                    The dependence of the UG predictions on the dataset weightings is shown for three different datasets. These datasets consider: all points included (left panel), an arbitrary pair removed (central panel), and the outliers removed (right panel). The TL energy estimates at 0-range are shown as a function of the dataset weighting $t$. In general, the estimates agree within a smaller margin when $t$ corresponds to the original dataset (emphasis on low $N$), but diverge when $t$ corresponds to the reflected dataset (emphasis on larger $N$ \emph{and} consequently the skewed slopes associated with the outliers). 
    			}\label{fig:t-dependence}
\end{center}
    		\end{figure}

    		To isolate the effect of the outliers, we compare three datasets that consider: all $N$, removing the outliers at $N=70$ and $N=80$, and removing an arbitrary pair (as a control). When the distribution favors low $N$ $(\text{near }t=1)$, Fig.~\ref{fig:t-dependence} shows that the predictions agree regardless of the fit used. However, as the distribution shifts more emphasis towards the outliers $(\text{near }t=-1)$, there is a significant deviation between different fits. This occurs very strongly in the all-in and arbitrarily removed datasets, but not when the outliers are removed. The rightmost panel removes the suspected outliers resulting in a significantly reduced variance in the estimates. By contrast, removing a similar but arbitrary pair (50, 90) does not have this effect. This suggests that this reduced variance is due to removing the outliers, and not just removing arbitrary points. This suggests two things: the inclusion of the outliers \emph{does} skew the predictions when emphasis is placed on them, and the original distribution $(t=1)$ doesn't provide a sufficient emphasis on the high $N$ to capture the associated effects. It is therefore appropriate to remove the outliers and use a uniform distribution for training.
    		% An alternate strategy could be to statistically impute them, however this implies we know the correct physics. 

    	\subsection{Removing Pairs}
            So far, we have only validated this effect against a single arbitrary pair being removed. To solidify this, we will generalize this procedure to multiple other pairs. For a fixed $F$, the datasets have three parameters: the linear fit form, the removed pair, and the balance parameter $t$. To measure the variation in predictions due to the different fits $s$, we take the difference between the maximum and minimum predictions for a network trained with a given pair $\mathcal{N}$ removed,
    		\begin{equation}
                s(\mathcal{N}) = \max_{t, f}(\xi_f(\mathcal{N}, t)) - \min_{t, f}(\xi_f(\mathcal{N}, t)),
            \end{equation}
            where $\xi_f(\mathcal{N}, t)$ is an energy prediction, $t\in[-1, 1]$, and $f\in\{\text{individual}, \text{overall}, \text{average}\}$ is the fit form used.

    		Unfortunately, this would require a large number of networks to be trained. To simplify this, we turn to Fig. 2 to construct a surrogate function which is easier to evaluate, but still captures the spread. Since the energy predictions tend to increase steadily from $t=1$ to $t=-1$ we evaluate only at these anchor points, which loosely captures half of the total spread, now defined as:
    		\begin{equation}
    			s(\mathcal{N}) \equiv \max_{f}(\xi_f(\mathcal{N}, -1)) - \min_{f}(\xi_f(\mathcal{N}, +1)).
    		\end{equation}
    		This only requires 6 networks to be trained per removed pair and is shown in Fig.~\ref{fig:t-dependence}. 
    		
    		    		\begin{figure}
    			\centering
\begin{center}
\includegraphics[width=1.0\columnwidth,clip=]
    			{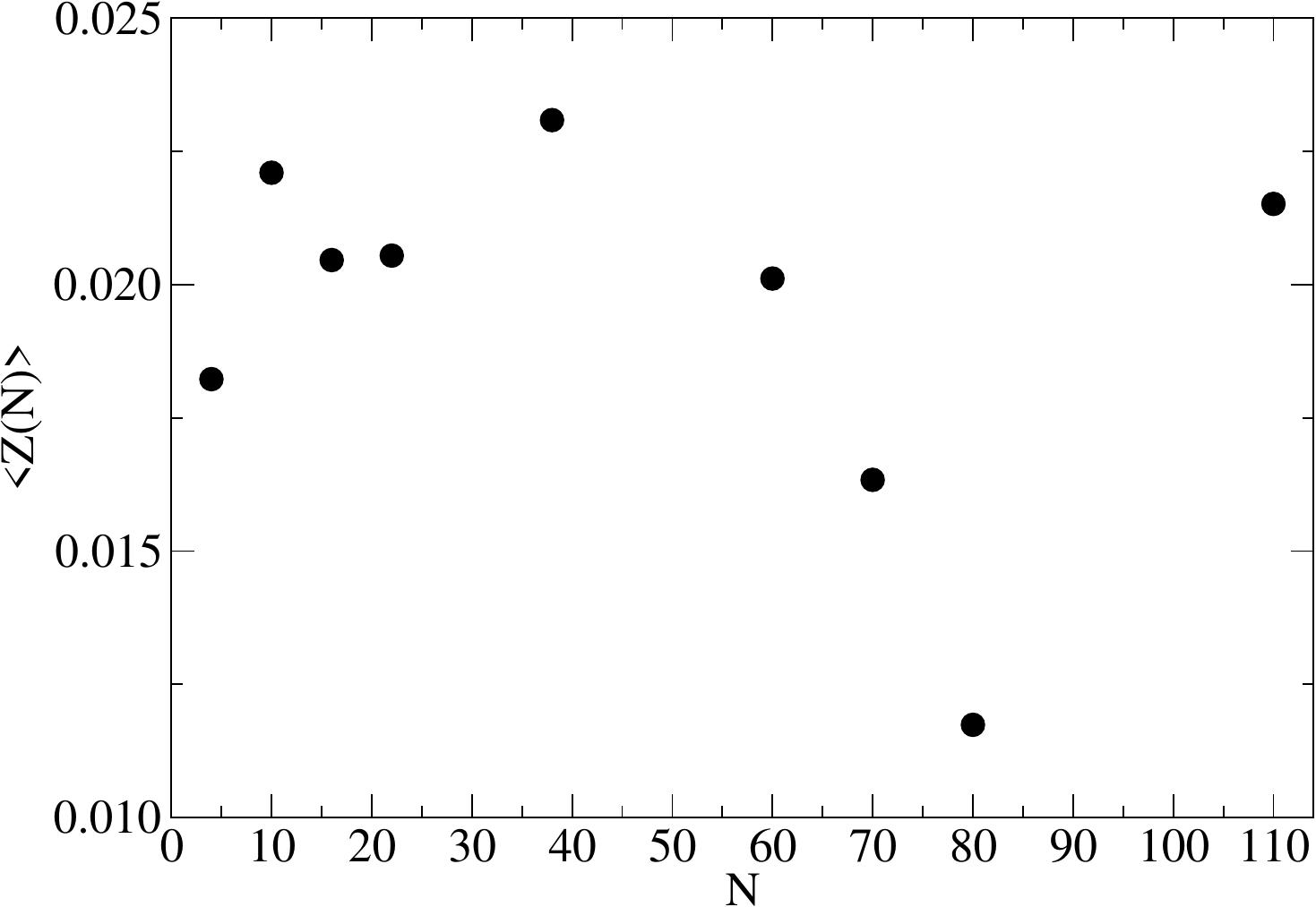}
    			\caption{
                    To verify that the pair of outliers identified in Fig.~\ref{fig:t-dependence} has a unique effect we study the removal of different pairs from the dataset. Networks are trained on different datasets, each with an arbitrary pair removed. The spread is averaged over predictions for pairs containing $N$, which produces the values shown in the figure, according to Eq.~(\ref{eq:spread}). There is a significant reduction at both $N=70$ and $N=80$, which suggests that the outliers do uniquely skew the predictions.
    			}
    			\label{fig:spread}
\end{center}
    		\end{figure}

            To estimate the spread associated with a given particle number (instead of a pair), one of the particle numbers in the pair is fixed to $N$. Then, the spread is averaged over a set of values for the 
            other particle numbers $N^*$,
    		\begin{equation}\label{eq:spread}
    			Z(N) \propto \sum_{N^*} s(\{N, N^*\}).
    		\end{equation}
    		The influence on the spread for each particle number is shown in Fig.~\ref{fig:spread}. At $N=70$ and $N=80$ the spread is significantly reduced. Thus we can claim that specifically removing the outliers has a unique effect and so they are  removed.

    	\subsection{Hyperparameter Optimization \& Predictions}\label{sec:UG_hyperparameter_optimization}
    		Having isolated and controlled small sample effects, we are now capable of carrying out a hyperparameter optimization. The individual fit will be used to augment the dataset since the other linear fits have served their purpose of identifying the outliers. The hyperparameters under consideration are: training dataset size, number of epochs, and hidden layer size. (A single ``epoch'' is when the whole dataset has been used once; it is standard to randomly select individual samples from the entire dataset many times over, i.e., for multiple epochs.) The training dataset size corresponds to $F$, noting that the actual count $C_N^F(t)$ gets rounded up according to the ceiling in Eq.~(\ref{eq:C_N^F}), and is then further split into a training and testing dataset. 
        
            % Dropout is a regularization technique aimed at reducing overfitting \textcolor{red}{[]}. The associated value is the rate that hidden layer units will be set to 0 during training. 
    		%http://www.jmlr.org/papers/volume15/srivastava14a/srivastava14a.pdf

    		                		\begin{figure}
\begin{center}
\includegraphics[width=1.0\columnwidth,clip=]
    		    {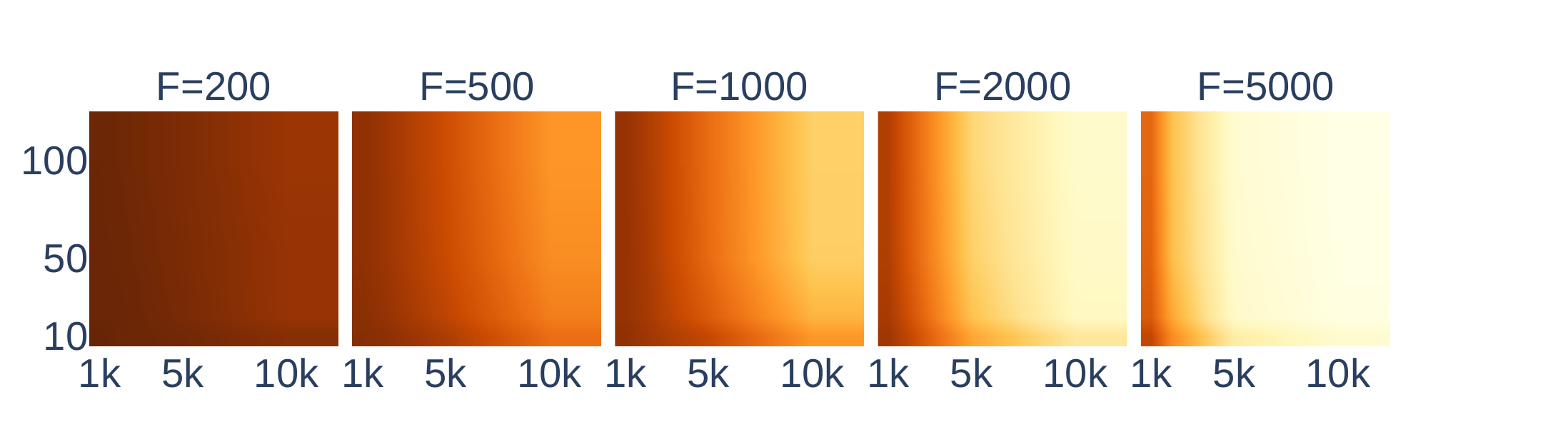}
    			\caption{
    			  The relative 5-fold cross-validation error for various hyperparameters is shown, where darker color indicates poorer performance. There are three parameters under consideration: each subplot contains a different dataset size $F$, with the number of training epochs along the bottom, and the number of hidden units along the vertical. In all cases, the performance stops improving after about 50 hidden units. For dataset sizes that are sufficiently large, the performance plateaus after about 5,000 epochs. The consistent improvements resulting from increasing the dataset size has a confounding effect which is discussed in the main text.
    			}
    			\label{fig:optimization}
\end{center}
    		\end{figure}

\begin{figure}[b]
\begin{center}
\includegraphics[width=1.0\columnwidth,clip=]
    			{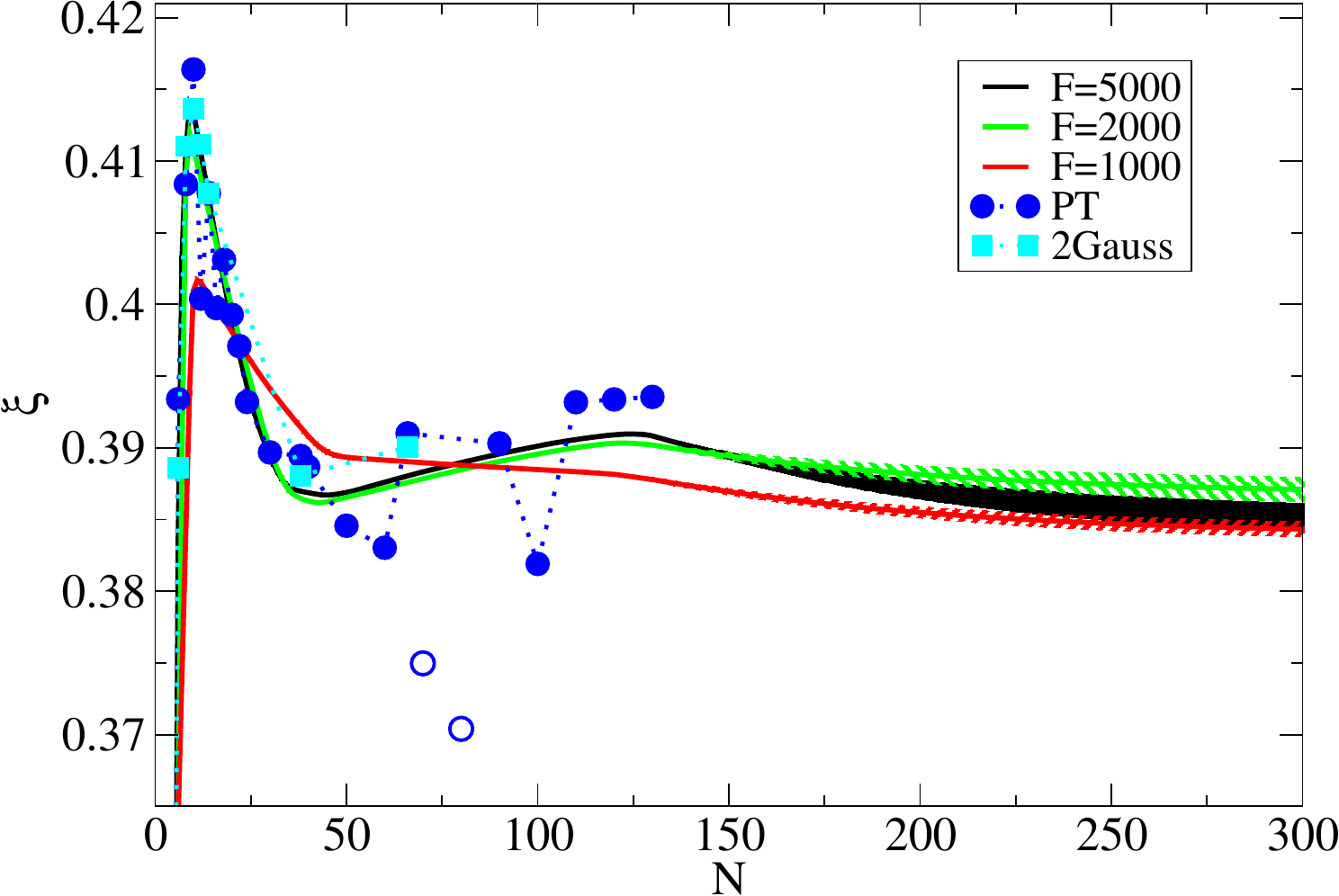}
    			\caption{
    			  A set of predictions from optimized neural networks (solid bands) trained on different dataset sizes, $F$, ordered in the caption the same way as in the plot itself at $N=100$. These are plotted alongside the original dataset (dotted) which is linearly extrapolated to 0-range. The outliers at $N=70$ and $N=80$ are hollow to denote that they were removed, as discussed in Section~\ref{sec:UG_hyperparameter_optimization}. The kink at $N\approx140$ in the black curve (solid) is evidence of overfitting, meanwhile the inability of the $F=1000$ curve (forward slash) to capture the peak at low $N$ is evidence of underfitting. The prediction from $F=2000$ (backward slash) is thus selected as the optimal network.
    			}
    			\label{fig:optimized_N_dependence}
\end{center}
    		\end{figure}

    		Performing the grid search depicted in Fig.~\ref{fig:optimization}, we find that epochs between 5,000 and 10,000, and hidden units around 50 have saturated performance.
            % Dropout doesn't seem to have any notable impact.
    		The training size requires further investigation since it has confounding effects. Ultimately, as the number of interpolated points increases arbitrarily, the training and testing set end up being strongly correlated. This means that increasing the value of $F$ beyond a certain point no longer provides an independent test. 		
    		With the other hyperparameters fixed, we can perform TL, 0-range predictions while varying the dataset size to identify signs of overfitting. This is done in Fig.~\ref{fig:optimized_N_dependence} which identifies overfitting for $F\ge5000$, and underfitting for $F\le1000$. This leaves the optimal network to have $F\approx2000$.

            % $\xi=0.389 \pm 0.001$
            The final step is to extract the TL, 0-range prediction from our optimal networks, which works out to be $\xi=0.389(1)$. There have been many calculations for the UG due to its great importance to many-body physics. Calculations are generally carried out as upper-bound estimates which have steadily decreased over time \cite{Carlson_2011, Forbes_2011, Gandolfi_2011}. Reference~\cite{forbes2012effectiverange}, whose raw data we employed as input for the present study, used simpler extrapolation techniques to arrive at the value $0.3897(4)$.
            The current leading theoretical value made use of a very large lattice calculation to arrive at $0.372(5)$ \cite{Carlson_2011, Carlson_2012}. Earlier experiments predicted values above $0.39$ \cite{Luo_2008, Navon_2010}, but have recently been improved to $0.376(4)$ \cite{Ku_2012}. Through additional considerations for the uncertainty involved, this estimate was lowered even further to $0.370(5)(8)$ \cite{Z_rn_2013}, in close agreement with the theoretical value. These calculations tend to predict lower energies than these network predictions, but this is not unexpected, given that DMC obeys a variational (i.e., upper bound) property. The networks do well to interpolate the provided dataset, which provides upper bound estimates on the true energy.

    \section{Learning the Effective Mass}
        \subsection{Dataset}
            The effective mass arises in Landau Fermi liquid theory (LFLT) \cite{i72:coleman:2015} and can be directly related to observables like the specific heat of neutron matter \cite{PhysRevC.89.044302}. It can be extracted by studying the dispersion relation for an excited particle in NM, as done in Ref.~\cite{Buraczynski_2019} and Ref.~\cite{Buraczynski_2020}. The quasi-particle energy $\Delta E_{T L}^{\left(k_{T L}\right)}$ is related to the momentum $k_\text{TL}^2$ through the dispersion relation:
            \begin{equation}\label{eq:dispersion}
                \Delta E_{T L}^{\left(k_{T L}\right)}=\frac{\hbar^{2} k_{T L}^{2}}{2 m^{*}},
            \end{equation}
            where $m^*$ is the effective mass. The subscripts ``TL'' refer to the notion that these quantities make use of an extrapolation prescription derived in Ref.~\cite{Buraczynski_2019}, which aims to reduce FSE. Despite this reduction, Eq.~(\ref{eq:dispersion}) still depends on the number of particles in our simulation, $N$, and so the FSE must still be studied.

\begin{figure}
                \centering
\begin{center}
\includegraphics[width=1.0\columnwidth,clip=]
                {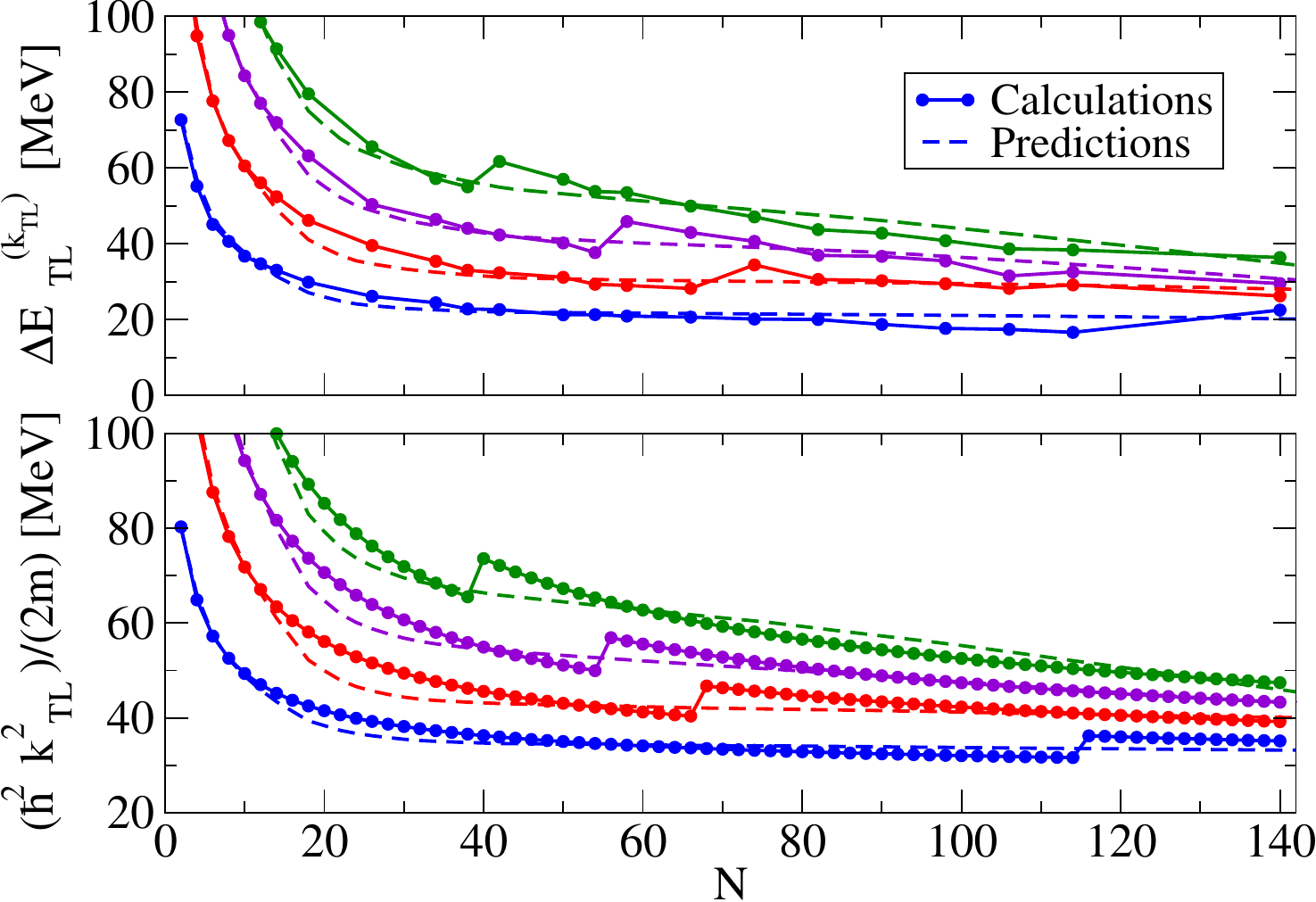}
                \caption{Interpolations predicted by the networks for the extrapolated quasi-particle energy $\Delta E_\text{TL}^{(k_\text{TL})}$ (top panel) and extrapolated momentum $k^2_\text{TL}$ (bottom panel) expressed in MeV, at a density of $0.05$ fm$^{-3}$. Both networks are trained on the same domain for consistency, although we show more points for the momentum since it is analytic. In both panels, the predictions (dashed) and calculated values (dots, solid) for the first four excited states are shown, increasing from bottom to top. Both extrapolated quantities have a similar dependence, which is generally captured by the networks. As expected, the discontinuities are not captured. However, since both models experience this same bias, the discontinuous predictions are resolved in the fit calculation for the effective mass.
                }
                \label{fig:similarity_factor}
\end{center}
            \end{figure}

            At a given $N$, we consider a system in its ground state and the associated energy, $E_N$. To probe the dispersion relation, we also consider adding a particle to this system in an excited state of momentum $k$ and the associated energy of the system, $E_{N+1}^{(k)}$. To access the TL behavior of these systems, the extrapolation prescription can be applied to both the energy and the momentum. The extrapolated momentum can be expressed in terms of single-particle states:
            \begin{align}
                k_\text{TL}^{2} &= k^2 - k^2_{F,N} + k_F^2 \\\nonumber
                &= \left( 2\pi\sqrt[3]{\frac{n}{N}} \right)^2 \bar{n}^2 - \left( 2\pi\sqrt[3]{\frac{n}{N}} \right)^2 \bar n^2_\circ + \left(3 \pi^{2} n\right)^{2 / 3} 
            \end{align}
            where $\bar n_\circ$ and $\bar n$ are the integer momenta for the ground state and excited state, respectively, not to be confused with the particle number density, $n$. Certain momentum values are inaccessible, due to the periodic boundary condition imposed in QMC, resulting in the discontinuities found in Fig.~\ref{fig:similarity_factor}. The quasi-particle energy is given in Ref.~\cite{Buraczynski_2020} as
            \begin{align}
                \Delta E_{N}^{(k)} = E_{N+1}^{(k)}-E_{N}+\frac{2}{5} \xi E_{F},
            \end{align}
            where $\xi$ is the Bertsch parameter, and $E_F$ is the Fermi energy. The extrapolated version of this, $\Delta E_{TL}^{\left(k_{TL}\right)}$ can be expressed in terms of the extrapolated momentum, a constant offset, and a potential energy term:
            \begin{align}\label{eq:deltaE_TL}
                \Delta E_{TL}^{\left(k_{TL}\right)} &= \Delta U_{N}^{(k)}+\frac{2}{5} (\xi - 1) E_{F} + \frac{\hbar^{2} k_{TL}^{2}}{2 m},
            \end{align}
            where $\Delta U_{N}^{(k)}$ is the difference in potential energy between both systems. The FSE introduced by this energy term are not large since, in neutron matter, the kinetic energy tends to be the dominant FSE contribution due to the small effective range of the interaction. In Fig.~\ref{fig:similarity_factor}, this is evident since the difference from the momentum is small. The discontinuities common to both the energy and momentum are due to the discretized nature of the available momentum states. 

            The effective mass is extracted from a linear fit between these quantities, according to Eq.~(\ref{eq:dispersion}). At a given density and particle number, the dispersion is studied by considering different excited states. To probe near the Fermi surface, as done in Ref.~\cite{Buraczynski_2020}, four excited states are used in the fit. The quasi-particle energy is an interacting quantity and is therefore limited by QMC to finite $N$. By contrast, the momentum is a non-interacting quantity and can be computed for arbitrarily large systems.

            In the ML task, the mapping to learn for the energy is:
            \begin{equation}
                (N, n) \xrightarrow{\text{Energy Networks}} \{\Delta E_{TL}^e\}_{e=1}^4,
                \label{eq:NNdef1}
            \end{equation}
            % Could have considered learning raw values, but this leaves additional work for the networks that we know for free.
            where the superscript $e$ labels the ordering of excited states, and $n$ again corresponds to the density.
            % Particles: Considerations include the upper and lower bounds, the stratification, and the inclusion of shell, off-shell, and shell-adjacent particle numbers. Ultimately many of those considerations were unnecessary and a dataset primarily comprised of shell values was sufficient for `medium' N.
            The machine learning task is to interpolate these calculations to extract the underlying particle-number dependency. This mapping naturally suggests a FFNN with 2 inputs and 4 outputs. Alternate schemes may consider breaking the mapping into individual excitations, however this would lose shared information between the different excited states. This notion of \textit{feature engineering}, whereby the structure of the inputs/outputs is selected with the aim of simplifying the model's learning task, is important to consider
            when performing extrapolations \cite{pastore2020extrapolating}.

            Since the momentum is a non-interacting quantity, it can be calculated to very large $N$. However, spurious effects are introduced in the fitting procedure if these are used. Networks tend to smooth the predictions and struggle with the discontinuities shown in Fig.~\ref{fig:similarity_factor}. 
            The training dataset for the quasi-particle energy contained evaluations for $N\in[1, 2, 3, 4, 5, 6, 7, 19, 27, 33, 45, 57, 70]$ and $n\in[0.02, 0.04, 0.06, 0.08, 0.1, 0.12, 0.14, 0.16, 0.18, 0.2]$. 
            During the fitting procedure, the mismatch between the energy and momentum continuities results in spurious predictions, as shown in Fig.~\ref{fig:discontinous_predictions_comparison}. 

            The solution here is to separately train networks to learn the momentum and join those results with the energy during the fit, as shown in Fig.~\ref{fig:discontinous_predictions_comparison}. For the momentum, the mapping to learn is
            \begin{equation}
                (N, n) \xrightarrow{\text{Momentum Networks}} \{k_{TL}^e\}_{e=1}^4.
                                \label{eq:NNdef2}
            \end{equation}
            Fitting Eqs.~(\ref{eq:NNdef1})~\&~(\ref{eq:NNdef2}) as per Eq.~(\ref{eq:dispersion}) provides an effective-mass prediction. Alternative formulations that attempted to resolve these discontinuities included the use of custom loss functions, and the use of transfer learning to initialize the networks with non-interacting trends. Ultimately, the momentum networks were required and sufficient. The hyperparameter optimization here is more difficult than in the UG case, since there are additional considerations for the two networks and the different features they may capture. For this, we will seek additional metrics.

        \subsection{Selecting Similar Networks}
            The discontinuous effective-mass predictions can be resolved by training networks to learn both the energy and momentum $N$-dependence, and combining the results in a fit. Since the energy and momentum networks are independent, different details may be captured between the two. While we expect the energy and momentum trends to differ according to Eq.~(\ref{eq:deltaE_TL}), the ability of the networks to capture these details may vary depending on the hyperparameters associated with each network. So, both network types are trained under a variety of hyperparameters, but in contrast to the UG, the testing error isn't a sufficient metric. Ultimately, the prediction quality depends on the combination of networks, whereas the testing error provides an independent metric. This can cause issues since two networks with similar testing errors may perform better/worse at different regions, resulting in a fit that could potentially be poor along the whole domain. To combat this, we aim to construct a selection criterion that identifies networks that capture similar features.

            \begin{figure}
                \centering
                \begin{center}
                \includegraphics[width=1.0\columnwidth,clip=]{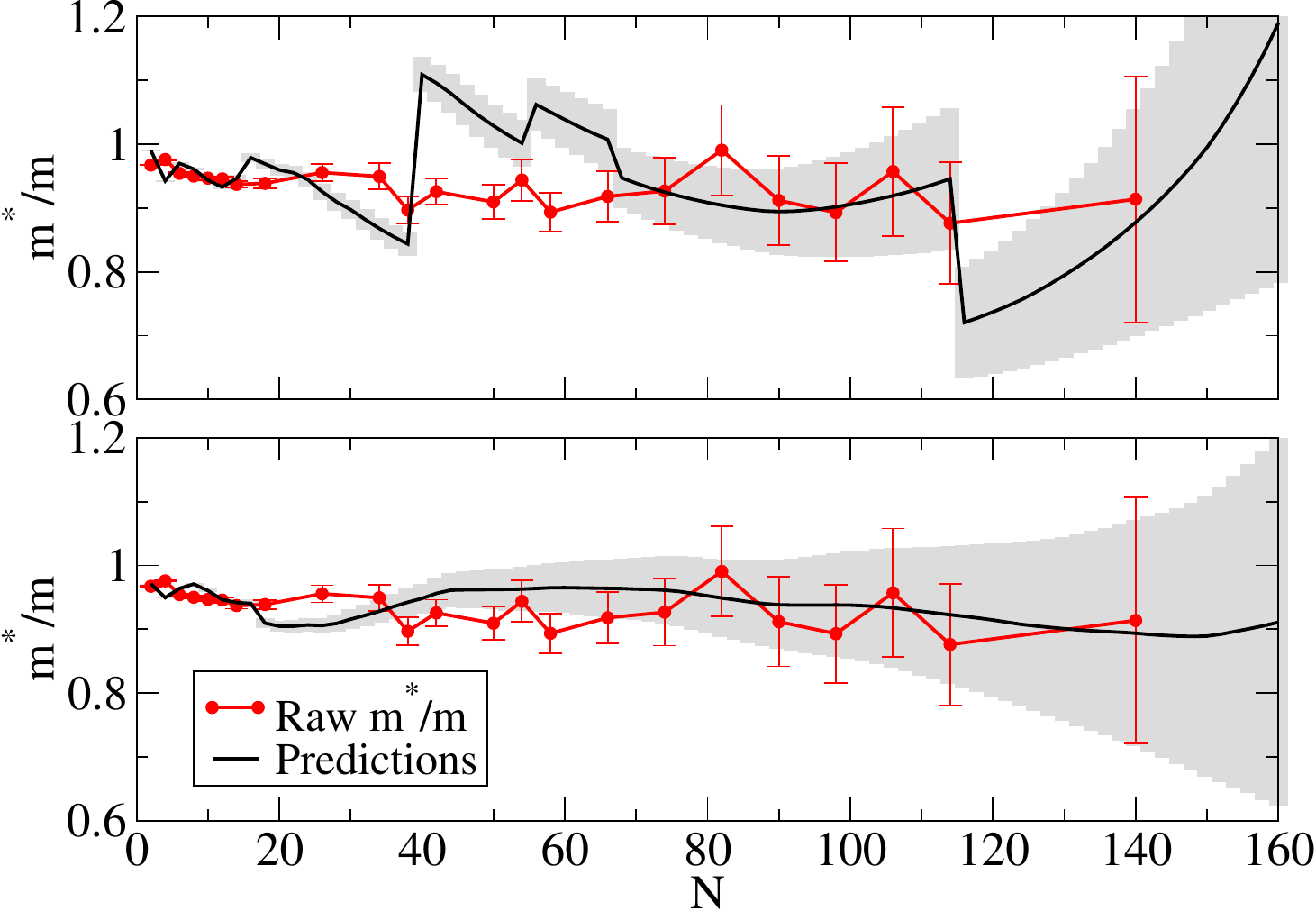}
                    \caption{
                        The effective masses predicted by neural networks at a density of 0.05 fm$^{-3}$ are contrasted for two fit methods. These predictions constitute an interpolation, since this density was not included during training. In the top panel, the energy predictions are fit to the analytic momentum, which results in discontinuities. In the bottom panel, the energy predictions are fit with predictions from a separate network that was trained on the momentum, which resolves the discontinuities. The smoothing effects shared by both network predictions eliminate the discontinuities in the fits. At low particle numbers, the effective-mass predictions are more sensitive to the large energy/momentum values (shown in Fig.~\ref{fig:similarity_factor}) which results in more variability. At high $N$, the predictions begin struggling to extrapolate and must contend with the larger QMC error.
                    }
                    \label{fig:discontinous_predictions_comparison}
                \end{center}
            \end{figure}

\begin{figure}
                \centering
 \begin{center}
\includegraphics[width=1.0\columnwidth,clip=]
               {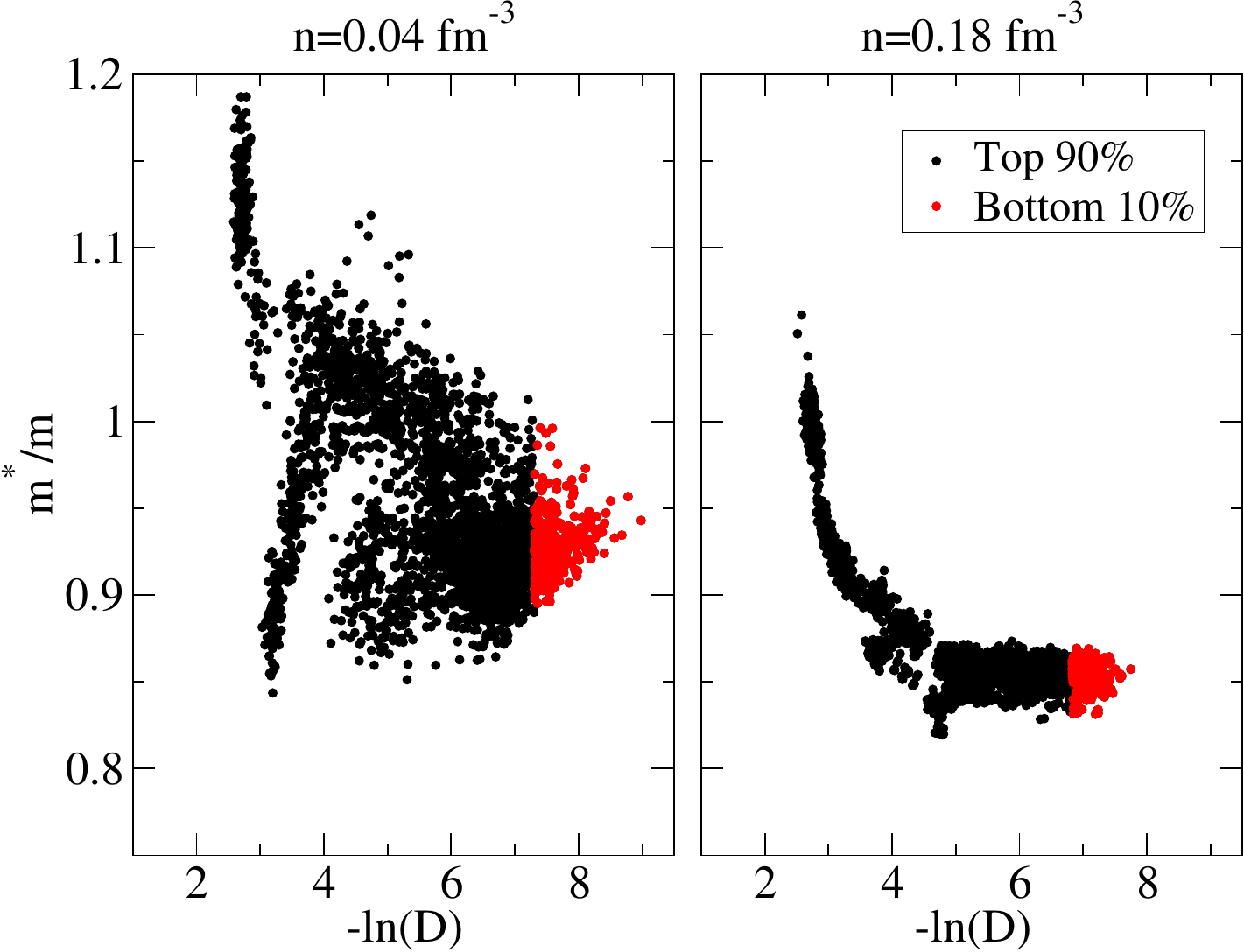}
                \caption{
                    The effective-mass predictions $m^*/m$ as a function of the disagreement factor $D$ for a density of 0.04 fm$^{-3}$ (left) and 0.18 fm$^{-3}$ (right) for networks with various hyperparameter settings. This contrasts the effect of increasing density; the networks at a lower density are more variable than those at higher densities. This is likely due to the reduced energy scale, which requires increased precision. Regardless of the density, there is a common trend that the network predictions limit towards some value as the disagreement factor decreases. At each density, the bottom 10\% (the rightmost portion) are selected to calculate an effective-mass statistic, as done for Fig.~\ref{fig:density_dependence}. These networks averaged their predictions over the domain $N\in[0, 130]$.
                }
                \label{fig:emr_vs_similarity_factor}
\end{center}
            \end{figure}

            The energy and momentum dependencies experience discontinuities due to the discrete momentum states, however the effective mass does not. In fact, the effective mass is relatively constant as shown in Fig.~\ref{fig:discontinous_predictions_comparison}. The claim that the effective mass \textit{is} constant would be too strong, however the effective-mass predictions should have a low variance across $N$. At a given density, we can evaluate the variance to define the \textit{disagreement factor} $D$ as a metric for the quality of a prediction: 
            \begin{align}
                D \equiv \text{Var} \left(\left\{\frac{m^*}{m}\left(n, N\right)\right\}_{N=N_\text{low}}^{N_\text{high}}\right),
            \end{align}
            where $n$ is the number density, and the variance is evaluated over particle numbers $N\in[N_\text{low}, N_\text{high}]$. This interval is used to exclude particle numbers where QMC and/or the networks may perform poorly. The predictions in the intermediate regions are generally stable, but may vary near the extremes. To identify possible boundary effects, we consider two domains: $N\in[0, 130]$, and $N\in[40, 120]$. The effective-mass predictions averaged over a given interval are used as a statistic. The effective-mass predictions from networks with lower disagreement factors have best captured the overall dependence and are therefore used to generate an estimate.  In general, the effective-mass predictions tend to converge to some value as the disagreement factor increases as shown in Fig.~\ref{fig:emr_vs_similarity_factor}. Each point on this plot corresponds to a pair of neural network ensembles and their associated set of hyperparameter values.
            The average prediction of networks with disagreement factors in the lowest 10\% is chosen as a statistic. At lower densities, the network predictions are less consistent. The networks may be struggling due to the difference in scale between higher and lower densities.

        \subsection{Predictions for Density Dependence}
\begin{figure}
                \centering
                \begin{center}
                    \includegraphics[width=1.0\columnwidth,clip=]{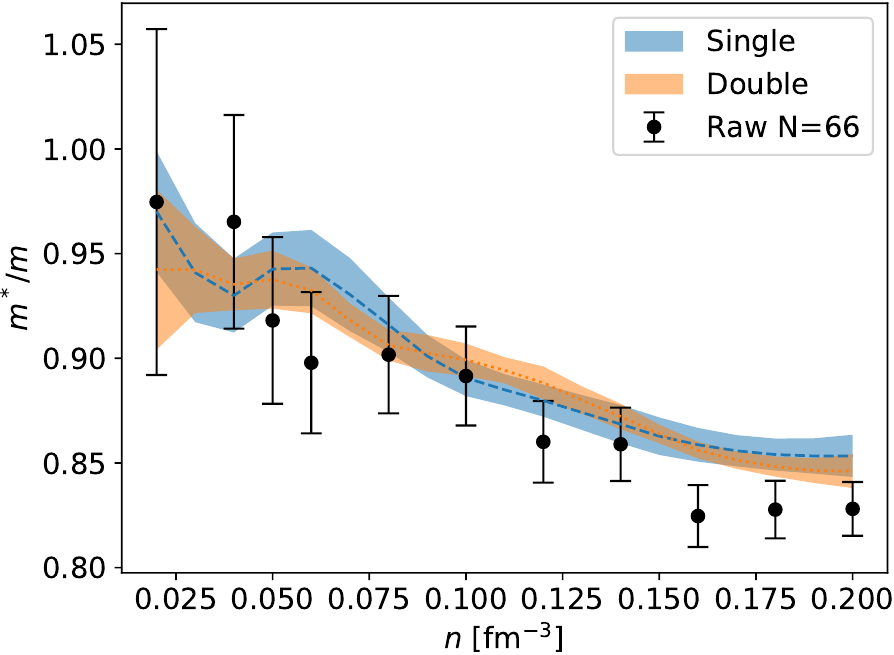}
                    \caption{
                        Network predictions for the effective mass $m^*/m$ as a function of number density $n$ along with raw AFDMC calculations for $N=66$. The blue dashed trend (which matches the raw calculation at the lowest density) shows predictions from networks with a single hidden layer, while the orange dotted trend uses networks with two equally-sized hidden layers. These predictions were averaged over the domain $N\in [0, 130]$. Overall, the agreement between the networks is quite good.
                    }
                    \label{fig:density_dependence}
                \end{center}
            \end{figure}

            We now have a systematic procedure for training neural networks to predict effective masses that are continuous. The use of both energy and momentum networks resolved the discontinuous predictions, and the disagreement factor was used to assess the effective-mass prediction quality. The predictions are averaged over a domain which is then used as an effective-mass statistic at a given density. This procedure can be repeated for a variety of densities as done in Fig.~\ref{fig:density_dependence}. We consider comparisons between different averaging domains and the effect of multiple hidden layers. In general, the agreement between all the predictions is quite good and they tend to agree within error.

            Predictions at the lowest densities experience the highest variance. Networks trained on the larger domain ($N\in[0, 130]$) tend to capture the overall effects with less uncertainty than the smaller domain. Although a network with a single hidden layer can in theory approximate a function as well as a multi-layered network, we would like to determine if multiple hidden layers may capture particular effects. To investigate this, we consider networks with two hidden layers of the same size. The predictions are similar, but tend to differ in subtle effects.

            As a comparison to previous work, the effective-mass predictions of Ref.~\cite{Buraczynski_2020} take the AFDMC calculation at $N=66$ as the estimate for each density. There is a good amount of overlap between the predictions and these raw estimates. In addition to the use of neural networks to carry-out this regression analysis, another significant contrast is that this work makes use of many particle numbers to provide a density estimate. For these reasons, the networks are expected to have captured additional details that would otherwise be missed. The networks are able to interpolate between densities and particle numbers and so effects like the slightly higher effective masses predicted by the networks at higher densities are expected to be better descriptions of the overall behavior. 
            % Ref.~\cite{Buraczynski_2020} also considers the density dependence of two nuclear potentials, which differ due to different forces captured. By contrast, the predictions here overlap since they stem from the same underlying potentials.

    \section{Summary \& Conclusion}
        We trained feed-forward neural networks on two related nuclear systems to study finite-size effects through machine learning techniques to understand potential issues, demonstrate techniques to resolve them, and improve predictions. The difficulty of carrying out high-precision calculations through QMC restricts our datasets to small sizes. We found the technique of data augmentation effective to mitigate these issues, although some care was required to account for introduced correlations. Despite increasing the number of data points available, the accuracy of the model still depends on the quality of the original dataset. Outliers can skew the data, but due to the scarcity of data, we take care to demonstrate their negative impact before justifying their removal. After validating the dataset, a hyperparameter optimization was carried out by using k-fold cross validation. The optimized networks were able to produce TL estimates for the zero-range limit of the UG, that were not trivial to access directly in a QMC approach.

        Networks can also struggle with other dataset effects, like discontinuities. They tend to predict smooth trends and struggle to capture sudden jumps. This smoothing bias introduces an inconsistency between the energy and momentum that was remedied by independently training networks to predict the associated momentum; as long as both quantities experienced similar behavior, the fitting process resulted in a continuous effective-mass prediction. The notion of \textit{network disagreement} was introduced to identify networks that satisfied this property. Following this, the networks were trained on data resulting from raw AFDMC energy calculations, which are given as supplementary material in Ref.~\cite{supple}. The TL density dependence of the effective mass for networks with single and double hidden layers were compared to raw calculations stemming from previous work.
        
        Overall, we have found that neural networks provide a versatile tool that allows one to capture finite-size effects for both ground-state and excited-state properties. This is especially important when there is no \textit{a priori} analytic expectation of what the $N$-dependence should be, as is often the case for strongly interacting systems. As seen in the results reported on in this work, it is possible to use machine learning to fold-in the entire $N$-dependence, without having to invoke simplifying approximations.

\begin{acknowledgments}
This work was supported by the Natural Sciences and Engineering Research Council (NSERC) of Canada, the
Canada Foundation for Innovation (CFI), and the Early
Researcher Award (ERA) program of the Ontario Ministry of Research, Innovation and Science. Computational resources were provided by SHARCNET and NERSC.
\end{acknowledgments}

    \bibliography{neural}

\end{document}